\documentclass[conference,a4paper]{IEEEtran}
\usepackage{xcolor}
\usepackage{balance}

\usepackage{cite}
\usepackage{amsmath,amssymb,amsfonts}
\usepackage{graphicx}
\usepackage{textcomp}
\usepackage{acronym}
\usepackage{xcolor}
\usepackage{tikz} 
\usepackage[utf8]{inputenc}
\usepackage{pgfplots} 
\usepackage{pgfgantt}
\usepackage{pdflscape}
\usepackage{changes}
\usepackage{comment}
\usepackage{subfigure}
\usepackage{mathtools,algpseudocode,algorithm,MnSymbol}
\usepackage{geometry}
\usepackage{pgfplots}
  \pgfplotsset{compat=newest}
  \usetikzlibrary{plotmarks}
  \usetikzlibrary{arrows.meta}
  \usepgfplotslibrary{patchplots}
  \usepackage{grffile}
  \usepackage{amsmath}

\pgfplotsset{compat=newest} 
\pgfplotsset{plot coordinates/math parser=false} 

\acrodef{GRAND}[GRAND]{guessing random additive noise decoding}
\acrodef{ML}[ML]{maximum likelihood}
\acrodef{ISAC}[ISAC]{integrated sensing and communication}
\acrodef{ORBGRAND}[ORBGRAND]{ordered reliability bits guessing random additive noise decoding}

\acrodef{UE}[UE]{user equipment}
\acrodef{EM}[EM]{electromagnetic}
\acrodef{ML}[ML]{maximum likelihood}
\acrodef{LS}[LS]{least-squares}
\acrodef{LMMSE}[LMMSE]{linear minimum mean square error}
\acrodef{MSE}[MSE]{mean square error}
\acrodef{BS}[BS]{base station}
\acrodef{VA}[VA]{virtual anchor}
\acrodef{SP}[SP]{scattering  point}
\acrodef{OFDM}[OFDM]{Orthogonal Frequency Division Multiplexing}
\acrodef{MIMO}[MIMO]{Multiple-Input Multiple-Output}
\acrodef{SISO}[SISO]{Single-Input Single-Output}
\acrodef{LLR}[LLR]{log-likelihood ratio}

\acrodef{LoS}[LoS]{line-of-sight}
\acrodef{NLoS}[NLoS]{non-line-of-sight}
\acrodef{TOA}[TOA]{time of arrival}
\acrodef{BLER}[BLER]{block error rate}

\acrodef{SISO}[SISO]{single-input single-output}

\acrodef{eBCH}[eBCH]{extended Bose–Chaudhuri–Hocquenghem}

\acrodef{ToA}[ToA]{time of arrival}
\acrodef{BPSK}[BPSK]{binary phase shift keying}
\acrodef{SNR}[SNR]{signal-to-noise Ratio}

\begin{document}

\bibliographystyle{IEEEtran}
\bstctlcite{IEEEexample:BSTcontrol}
%
    \title{Sensing-Aided Ordered Reliability Bits Guessing Random Additive Noise Decoding}

\author{\IEEEauthorblockN{
Yu Ge\IEEEauthorrefmark{1},   
Lukas Rapp\IEEEauthorrefmark{1}, Ken R. Duffy\IEEEauthorrefmark{2}, Muriel Médard\IEEEauthorrefmark{1}      
}                                     
\IEEEauthorblockA{\IEEEauthorrefmark{1}
Research Laboratory for Electronics, Massachusetts Institute of Technology, Cambridge, USA,\\
\IEEEauthorrefmark{2}
College of Engineering and College of Science, Massachusetts Northeastern University, Boston, USA,\\}
\IEEEauthorblockA{ 
\{yuge,~rappl,~medard\}@mit.edu, k.duffy@northeastern.edu}}



\maketitle

\begin{abstract}
\Ac{ISAC} is a key enabler for future wireless systems, providing environmental information that can support tasks beyond conventional data transmission. However, its impact on channel decoding remains less explored. This paper studies sensing-aided \ac{ORBGRAND} over single-input single-output narrowband fading channels. Environmental information is used to construct a geometry-based prior for the channel coefficient, which is fused with pilot observations via \ac{LMMSE} estimation. The resulting posterior channel estimate and uncertainty are used to compute the \acp{LLR} supplied to \ac{ORBGRAND}, improving the reliability ordering that drives its noise-guessing process. Simulation results demonstrate improved block error rate and reduced average query complexity, with the largest gains in pilot-limited regimes.
\end{abstract} 
\vskip0.5\baselineskip

\section{Introduction}

\Acf{ISAC} has emerged as a key technology for future wireless networks, where communication systems are expected not only to transmit data but also to sense and understand the surrounding environment \cite{An_ISAC_Survey2022,procIEEE_2023_ISAC_6G}. Radio sensing lies at the heart of \ac{ISAC}, where ISAC systems can infer geometric information, such as the states of connected devices and passive objects in the propagation environment, by exploiting signals that propagate through and interact with the environment \cite{ge2023mmwave,ge2025sensing}. Owing to its broad range of potential applications, radio sensing has attracted considerable research attention \cite{Amjad2023,ge2023mmwave}, and recent real-world experiments have demonstrated its practical feasibility \cite{ge2024experimental,rastorgueva2024millimeter}.

Beyond localization and mapping, sensing information can also support physical-layer communication. Many existing ISAC studies use environmental information to improve physical-layer transmission before decoding, for example, by refining channel estimates, predicting beam directions, or reducing pilot overhead \cite{liu2022integrated,wymeersch2025cross}. However, its direct impact on channel decoding has received much less attention. In conventional soft-decision decoding architectures, channel estimation and demodulation are performed before decoding, producing bit-wise \acfp{LLR} that are then consumed by the channel decoder \cite{caire1998bit,martinez2009bit}. This separation overlooks an important opportunity: sensing can improve the reliability of the soft information supplied to the decoder, and therefore should be incorporated into the decoding process.

\Acf{GRAND} is a recently developed universal decoding framework for channel codes \cite{duffy2019capacity,galligan2021igrand,riaz2021multi}. Instead of designing a decoder around a specific code structure, GRAND decodes by guessing the binary noise effects that corrupted the transmitted codeword and checking whether the resulting candidate satisfies the code constraints \cite{duffy2019capacity}. \Acf{ORBGRAND} is a practical soft-decision variant of GRAND that exploits reliability information from the received symbols. Specifically, ORBGRAND uses the ordering of bit reliabilities to construct a near-likelihood-ordered sequence of noise effects, so that less reliable bit positions are flipped earlier \cite{duffy2022ordered,liu2022orbgrand,an2022keep,riaz2023sub}. Since ORBGRAND relies directly on the rank ordering of the magnitudes of the input \acp{LLR}, its decoding performance and query complexity are contingent on channel-estimation quality.

This dependence creates a natural connection between ISAC and ORBGRAND: sensing information can improve channel estimation, which in turn improves the \acp{LLR} and the noise-query order used by ORBGRAND. Recent works have investigated the impact of channel-estimation errors in GRAND-based receivers \cite{duffy2023using,wiame2025joint}. However, to the best of our knowledge, the use of environmental sensing information to assist GRAND-based frameworks has not yet been studied. In this paper, we investigate sensing-aided ORBGRAND, where sensing improves decoding through the soft-information pipeline. Specifically, sensing information is used to construct a geometry-based prior for the channel coefficient, which is fused with pilot observations through \acf{LMMSE} channel estimation, and the resulting channel estimate and uncertainty are used to compute the \acp{LLR} supplied to ORBGRAND. 

The main contributions of this paper are summarized as follows: \textit{(i)} We connect ISAC with GRAND-based frameworks by showing how environmental side information can improve ORBGRAND reliability ordering through channel-estimation priors; \textit{(ii)} We develop a geometry-based channel prior using sensing information and fuse it with pilot observations via LMMSE estimation to generate sensing-aided \acp{LLR} for ORBGRAND; \textit{(iii)} We validate the effectiveness of the proposed framework, demonstrating the performance gain brought by sensing information within the decoding pipeline.

\subsubsection*{Notation} Scalars (e.g., $x$) are denoted in italic, and vectors (e.g., $\boldsymbol{x}$) in bold, matrices (e.g., $\boldsymbol{X}$) in bold capital letters. The transpose is denoted by $(\cdot)^{\top}$. The Hermitian transpose is denoted by $(\cdot)^{\text{H}}$. The L2 norm is denoted by $||\cdot||$. The $i$-th component in vector $\boldsymbol{x}$  is denoted by $\boldsymbol{x}_{i}$.

\section{Signal and system models}
In this paper, we consider a pilot-aided \acf{SISO} scenario, in which the transmitter sends narrowband signals to the receiver through the propagation environment. This section introduces the received signal model and the system model.

\subsection{End-to-End System Model}
We consider a pilot-aided coded communication system, in which environmental side information is available at the receiver. A binary channel encoder maps an information message to a length-$n$ binary codeword $\boldsymbol{c}=[c_1,\ldots,c_n]^{\top} \in \{0,1\}^n$, which belongs to a binary linear block code $\mathcal{C}$ with parity-check matrix $\boldsymbol{H}$. The coded bits are then mapped to the \ac{BPSK} symbols according to
\begin{equation}
    x_i = 1 - 2c_i, \qquad i=1,\ldots,n. \label{BPSK}
\end{equation}

Before data transmission, the transmitter sends known pilot symbols for channel estimation. Depending on the available side information, the receiver either estimates the channel coefficient using pilots only, or fuses the pilot observations with environmental information to obtain an improved channel estimate and an associated uncertainty. The resulting channel estimate is then used to compute bit-wise \acp{LLR} for the received coded symbols. Finally, the \ac{LLR} vector and the parity-check matrix $\boldsymbol{H}$ are provided to ORBGRAND, which returns an estimated codeword $\hat{\boldsymbol{c}}\in\mathcal{C}$. Environmental information does not modify the code or  ORBGRAND directly. Instead, it affects the channel estimate $\hat{h}$ and the resulting LLR vector $\boldsymbol{L}$ supplied to the decoder.

\subsection{Signal Models} \label{Signal model}
In the considered scenario, the transmitter sends signals to the receiver, which can reach the receiver  directly, termed as the \ac{LoS} path, and/or bounced off by the landmarks in the environment, for example, reflected by the reflecting surface, and then reach the receiver, termed as \ac{NLoS} paths. We work with an equivalent discrete-time complex baseband model obtained after pulse shaping at the transmitter and matched filtering and symbol-rate sampling at the receiver {\cite[Chapter~3]{sklar2021digital}}. Under this model, when the narrowband signals are sent, the received signal is given by
\begin{equation}
    y_{i} = h x_{i} + w_{i}, \qquad i=1,\ldots,n, \label{channel}
\end{equation}
where $x_{i}$ denotes the $i$-th transmitted symbol, $y_{i} \in \mathbb{C}$ is the corresponding received symbol, $w_i \sim \mathcal{CN}(0,\sigma_w^2)$ is complex circularly symmetric white Gaussian noise with known variance, where $\sigma_w^2$ denotes the noise variance per received symbol, and $h\in\mathbb{C}$ is the narrowband channel coefficient. We assume that the channel $h$ does not change during transmission of $n$ symbols. Since the channel consists of multipath components, and we explicitly model one \ac{LoS} path and one \ac{NLoS} path reflected by a reflecting surface, the channel coefficient can be expressed as 
\begin{align}
    h & = \underbrace{\rho_{\text{LoS}}  e^ {-\jmath 2 \pi f_{\mathrm{c}} \tau_{\text{LoS}}}}_{h_{\text{LoS}}} + \underbrace{\rho_{\text{NLoS}}  e^ {-\jmath 2 \pi f_{\mathrm{c}} \tau_{\text{NLoS}}}}_{h_{\text{NLoS}}} + h_{\text{res}},\label{eq:channel}
\end{align}
where $f_{\mathrm{c}}$ is the carrier frequency, $h_{\text{LoS}}$ is the LoS channel component, $h_{\text{NLoS}}$ is the reflected-path channel component,  $h_{\text{res}}\sim \mathcal{CN}(0,\sigma^2_{\text{res}})$ denotes the channel effect caused by other residual unmodeled scattering, modeling mismatch, and other effects not captured by the two-path model, with $\sigma^2_{\text{res}}$ being its variance. Moreover, $\rho_{\text{LoS}}$ and $\tau_{\text{LoS}}$ are the channel gain and the \ac{ToA} of the LoS path, respectively, and $\rho_{\text{NLoS}}$ and $\tau_{\text{NLoS}}$ are the channel gain and the \ac{ToA} of the reflected path, respectively. 

The LoS and the reflected-path channel components are determined by the channel gain and the \ac{ToA} of each path, which are given  by
\begin{align}
    \rho_{\text{LoS}}&=\frac{\lambda}{4\pi\left\|\boldsymbol{x}_{\mathrm{Tx}}-\boldsymbol{x}_{\mathrm{Rx}}\right\|}\label{gain_los}\\
\tau_{\text{LoS}}&=\frac{\left\|\boldsymbol{x}_{\mathrm{Tx}}-\boldsymbol{x}_{\mathrm{Rx}}\right\|}{c}\label{delay_los}\\
 \rho_{\text{NLoS}}&=\frac{\lambda\Gamma e^{\jmath \pi}}{4\pi\left\|\boldsymbol{x}_{\mathrm{VA}}-\boldsymbol{x}_{\mathrm{Rx}}\right\|}\label{gain_nlos}\\
\tau_{\text{NLoS}}&=\frac{\left\|\boldsymbol{x}_{\mathrm{VA}}-\boldsymbol{x}_{\mathrm{Rx}}\right\|}{c}\label{delay_nlos}
\end{align}
where $\lambda$ is the wavelength, $c$ is the speed of light, and $\Gamma \in [0,1]$ denotes the reflection attenuation factor of the reflecting surface. We also assume a constant phase shift of $\pi$ for the reflected path, corresponding to the ideal specular reflection. In addition, $\boldsymbol{x}_{\mathrm{Tx}}$ denotes the position of the transmitter, which is known, $\boldsymbol{x}_{\mathrm{Rx}}$ denotes the position of the receiver, and $\boldsymbol{x}_{\mathrm{VA}}$ denotes the location of the \ac{VA} of the reflecting surface, which is the reflection of the transmitter with respect to the reflecting surface, serving as a compact representation of the surface and is surface-specific \cite{palacios2019single}. In this paper, we assume that the sensing task has already been performed, and that the receiver position $\boldsymbol{x}_{\mathrm{Rx}}$ and the VA location $\boldsymbol{x}_{\mathrm{VA}}$ are available at the receiver.

\section{GRAND and ORBGRAND}

In this section, we introduce the basics of the GRAND and ORBGRAND for decoding a binary linear code, and describe how the \acp{LLR} are computed.

\subsection{GRAND as Noise Guessing}

For the parity-check matrix $\boldsymbol{H}$, the transmitted binary codeword
$\boldsymbol{c}$ satisfies
\begin{equation}
    \boldsymbol{H}\boldsymbol{c}=\boldsymbol{0}\mod 2 .
\end{equation}
Given the received soft information, the decoder first forms a hard-decision vector
$\boldsymbol{z}\in\{0,1\}^n$. GRAND then queries binary noise effects
$\boldsymbol{e}\in\{0,1\}^n$ and forms the candidate \cite{duffy2019capacity}
\begin{equation}
    \hat{\boldsymbol{c}}
    =
    \boldsymbol{z}\oplus\boldsymbol{e},
\end{equation}
where $\oplus$ denotes bit-wise XOR. Each candidate is checked for codebook membership via the parity-check condition
\begin{equation}
    \boldsymbol{H}\hat{\boldsymbol{c}}
    =
    \boldsymbol{0}
    \mod 2 .
\end{equation}
The first candidate satisfying the parity-check constraints is returned as the decoded codeword. Therefore, the performance and complexity of GRAND depend on the order in which the candidate noise effects are queried \cite{duffy2019capacity}. If the true noise effect appears early in the query order, decoding succeeds with low complexity; otherwise, many queries may be required. In addition, under a finite query budget, decoding may be abandoned, or an incorrect codeword might be found earlier, resulting in decoding errors.

\subsection{ORBGRAND}

ORBGRAND is a soft-input GRAND variant that uses the magnitudes of the bit-wise \acp{LLR} to determine which bit positions are more likely to be erroneous \cite{duffy2022ordered}. Let $\boldsymbol{L}=[L_1,\ldots,L_n]^{\top}$ denotes the bit-wise \ac{LLR} vector supplied to the decoder, where
\begin{equation}
    L_i=\log\frac{p(c_i=0|y_i)
    }{p(c_i=1|y_i)}=\log\frac{p(y_i|c_i=0)
    }{p(y_i|c_i=1)}, \label{LLR}
\end{equation}
where the equality follows from the assumption of equiprobable coded bits. The hard decision is obtained from the sign of the \ac{LLR} as
\begin{equation}
    z_i =\begin{cases}
        0, & L_i \geq 0,\\
        1, & L_i < 0,
    \end{cases} \qquad i=1,\ldots,n .
\end{equation}
The magnitude $|L_i|$ quantifies the reliability of this decision, where a small value of $|L_i|$ indicates that the receiver is uncertain about the corresponding bit, whereas a large value indicates a more reliable decision. ORBGRAND sorts the bit positions from least reliable to most reliable. Let $\kappa$ be a permutation satisfying
\begin{equation}
    |L_{\kappa(1)}|\leq|L_{\kappa(2)}|\leq \cdots \leq|L_{\kappa(n)}| .
\end{equation}
Noise effects involving less reliable bit positions, i.e., $\kappa(1)$, are queried earlier. Hence, when the \acp{LLR} accurately reflect the bit reliabilities, the correct noise effect is more likely to appear early in the query order.

In addition to the \ac{BLER}, query complexity is a key performance metric for ORBGRAND. We denote by $N_{\mathrm{guess}}$ the number of queried noise effects before the decoder terminates. In practical implementations, the decoder may also be constrained by a maximum query budget $T_{\max}$, in which case decoding is abandoned if no valid codeword is found within $T_{\max}$ queries. 

\subsection{\acp{LLR} under Imperfect Channel Knowledge}
The \ac{LLR} vector is the fundamental soft input to ORBGRAND \cite{duffy2022ordered}. It determines both the hard-decision vector used as the starting point of decoding and the reliability ordering used to construct the noise-query sequence. Therefore, the quality of the \acp{LLR} directly affects not only the decoding performance, but also the number of noise effects that ORBGRAND must query before termination. For the \ac{BPSK} mapping in \eqref{BPSK} and the received signal in \eqref{channel}, given a channel estimate $\hat{h}$, the receiver computes the \ac{LLR} for the $i$-th coded bit using \eqref{LLR}. Under the complex Gaussian noise model, this can be written as
\begin{equation}
    L_i =\log\frac{p(y_i|c_i=0,\hat{h})
    }{p(y_i|c_i=1,\hat{h})}\approx \frac{4\operatorname{Re}\{\hat{h}^{*}y_i\}}{\sigma_{\mathrm{eff}}^2
    },\label{eq:llr_imperfect_csi}
\end{equation}
where $\sigma_{\mathrm{eff}}^2$ is the effective noise variance used in the \ac{LLR} computation, and $\hat{h}$ is the channel estimate. Note that \eqref{eq:llr_imperfect_csi} is equivalent to marginalization over the posterior channel distribution \cite{wiame2025joint}. With perfect channel knowledge, $\hat{h}=h$ and {$\sigma_{\mathrm{eff}}^2=\sigma_w^2$}. With imperfect channel estimation, residual channel uncertainty can be incorporated by increasing the effective variance. It is clear in \eqref{eq:llr_imperfect_csi} that the \acp{LLR} are computed based on $\hat{h}$, and the channel estimate affects both the sign and magnitude of the \acp{LLR}. The sign determines the hard-decision vector $\boldsymbol{z}$, while the magnitude determines the reliability ordering used by ORBGRAND.

\section{Sensing-Aided Channel Estimation for ORBGRAND}
Environmental sensing information can improve channel estimation and the associated uncertainty quantification, which in turn determines the \acp{LLR} supplied to ORBGRAND. In this section, we describe how environmental sensing information is incorporated into channel estimation and how the resulting estimate is used to compute the \acp{LLR} for ORBGRAND. 

\subsection{Pilot-Only Channel Estimation} \label{Sec:pilot_only}

We first consider conventional pilot-only channel estimation. Before data transmission, the transmitter sends known pilot symbols that are used for channel estimation, and the received pilot signals are given by
\begin{equation}
    \boldsymbol{y}_{\text{p}}= h\boldsymbol{x}_{\text{p}}+\boldsymbol{w}_{\text{p}},\label{pilot_signal}
\end{equation}
where $\boldsymbol{x}_{\text{p}}\in\mathbb{C}^{N_\text{p}}$ is the known pilot sequence, with $N_\text{p}$ denoting its length, $\boldsymbol{y}_{\text{p}}\in\mathbb{C}^{N_\text{p}}$ is the corresponding received pilot sequence, and $\boldsymbol{w}_{\text{p}}\sim\mathcal{CN}(\boldsymbol{0},\sigma_w^2\mathbf{I})$ is complex-valued white Gaussian noise. We further denote
the pilot energy as $E_{\text{p}}=\boldsymbol{x}_{\text{p}}^{H}\boldsymbol{x}_{\text{p}}$. Given the known pilot sequence $\boldsymbol{x}_{\text{p}}$ and the received pilot sequence $\boldsymbol{y}_{\text{p}}$, we can perform channel estimation to estimate the channel $h$. The \ac{LS} channel estimate is given by
\begin{equation}
    \hat{h}_{\mathrm{LS}}=\frac{\boldsymbol{x}_{\text{p}}^{H}\boldsymbol{y}_{\text{p}}}{E_{\text{p}}}.\label{LS_estimation}
\end{equation}
Substituting \eqref{pilot_signal} into \eqref{LS_estimation}, we have 
\begin{equation}
    \hat{h}_{\mathrm{LS}}=h+\frac{\boldsymbol{x}_{\text{p}}^{H}\boldsymbol{w}_{\text{p}}
    }{E_{\text{p}}}=
    h + \tilde{w}_{\mathrm{LS}},
\end{equation}
where the estimation error follows $\tilde{w}_{\mathrm{LS}}\sim \mathcal{CN}\left(0,\sigma_{\mathrm{LS}}^2\right)$. Thus, the \ac{LS} estimate is unbiased, and its error variance is
\begin{equation}
    \sigma_{\mathrm{LS}}^2
    =
    \mathbb{E}\left[
        |\hat{h}_{\mathrm{LS}}-h|^2
    \right]
    =
    \frac{\sigma_w^2}{E_{\text{p}}}.
\end{equation}
For unit-power pilots, $E_{\text{p}}=N_{\text{p}}$, and the uncertainty of pilot-only estimation decreases with the pilot budget. When the pilot length is small, however, the channel estimate may be inaccurate, leading to unreliable \acp{LLR} for ORBGRAND.

\subsection{Channel Prior from Sensing Information}

Environmental sensing information can provide geometric knowledge about the propagation channel. Let $\mathcal{I}_{\mathrm{sen}}$ denote the available sensing information, such as the receiver position $\boldsymbol{x}_{\mathrm{Rx}}$, the \ac{VA} location $\boldsymbol{x}_{\mathrm{VA}}$, and the reflection attenuation factor $\Gamma$ associated with a reflecting surface. Based on the two-path model in Section~II, this information can be used to predict a part of the channel coefficient. We represent this prediction via a conditional Gaussian prior
\begin{equation}
    p(h|\mathcal{I}_{\mathrm{sen}})
    \sim
    \mathcal{CN}
    \left(
        \mu_{\mathrm{sen}},
        \sigma_{\mathrm{sen}}^2
    \right),\label{channel_prior}
\end{equation}
where $\mu_{\mathrm{sen}}$ is the sensing-predicted channel mean and $\sigma_{\mathrm{sen}}^2$ captures the remaining channel uncertainty. Different levels of sensing information lead to different priors, and we consider the following cases:

\begin{itemize}
\item  \textbf{Rx-aided}: If only the receiver position is available, the receiver can predict the \ac{LoS} component, so that $\mu_{\mathrm{sen}}=h_{\mathrm{LoS}}$ according to \eqref{gain_los}--\eqref{delay_los}.
The reflected path and residual propagation effects remain uncertain. In this case, the prior variance include the uncertainty due to the unmodeled reflected component and residual scattering,  i.e., $\sigma_{\mathrm{sen}}^2=\sigma_{\mathrm{NLoS}}^2+\sigma_{\mathrm{res}}^2$, where $\sigma_{\mathrm{NLoS}}^2$ and $\sigma_{\mathrm{res}}^2$ denote the uncertainties of the unmodeled reflected component and residual scattering, respectively.
\item  \textbf{Rx+VA-aided}:
If both the receiver and the \ac{VA} information are available, the receiver can predict both the \ac{LoS} and reflected components, so that $\mu_{\mathrm{sen}}=h_{\mathrm{LoS}}+h_{\mathrm{NLoS}}$ according to \eqref{gain_los}--\eqref{delay_nlos}. The corresponding prior variance is smaller, since only residual modeling errors and unmodeled scattering remain, i.e., $\sigma_{\mathrm{sen}}^2=\sigma_{\mathrm{res}}^2$. Therefore, sensing information improves channel estimation by providing a more accurate prior mean and reducing the prior variance.
\item  \textbf{Mismatched prior}: If there are  errors in the receiver and/or the \ac{VA} information, the predicted channel mean may be biased, while the corresponding prior variance remains small. Therefore, inaccurate sensing information can yield a biased and overconfident channel  prior.
\end{itemize}

\subsection{Sensing-Aided LMMSE Estimation}

The sensing prior can be fused with the pilot observation through scalar \ac{LMMSE} estimation. Given the \ac{LS} channel estimate in \eqref{LS_estimation} and the conditional Gaussian prior for the channel in \eqref{channel_prior}, the LMMSE estimate is \cite[Chapter~8.3]{oppenheim2017signals}
\begin{equation}
    \hat{h}_{\mathrm{sen}}=\mu_{\mathrm{sen}}+\frac{\sigma_{\mathrm{sen}}^2}{\sigma_{\mathrm{sen}}^2+\sigma_{\mathrm{LS}}^2}
    \left(\hat{h}_{\mathrm{LS}}-\mu_{\mathrm{sen}}\right).
    \label{eq:sensing_lmmse}
\end{equation}
The corresponding posterior variance is
\begin{equation}
    \sigma_{\mathrm{post}}^2=\frac{\sigma_{\mathrm{sen}}^2 \sigma_{\mathrm{LS}}^2}{\sigma_{\mathrm{sen}}^2+\sigma_{\mathrm{LS}}^2}.
    \label{eq:posterior_variance}
\end{equation}

\begin{figure}
\center
\centerline{\includegraphics[width=1\linewidth]{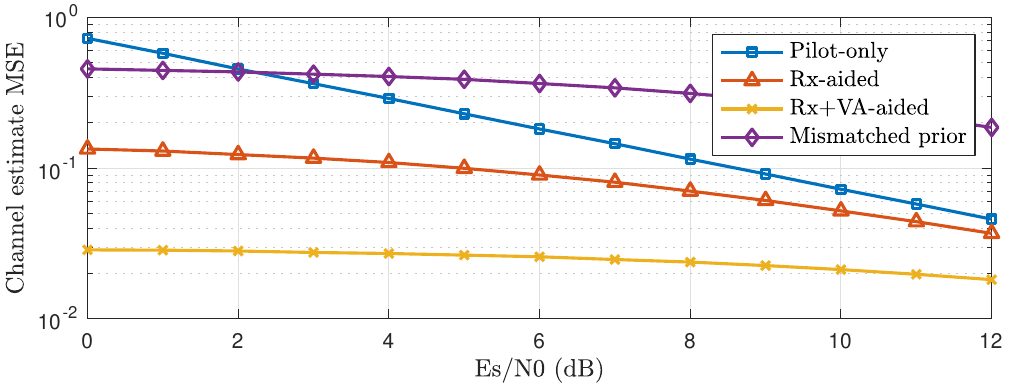}}
\vspace{-4mm}
\caption{Channel estimation MSE versus the energy per symbol to noise power spectral density ratio ($E_s/N_0$) for pilot-only, Rx-aided, Rx+VA-aided, and mismatched-prior (accurate Rx, and inaccurate VA) LMMSE estimation ($N_p=2$).
}
\label{fig:CE}
\end{figure}

From \eqref{eq:sensing_lmmse}, we can observe that when the pilot estimate is reliable, $\sigma_{\mathrm{LS}}^2$ is small and the estimator relies mainly on $\hat{h}_{\mathrm{LS}}$. When the pilot budget is small or the pilot observation is noisy, $\sigma_{\mathrm{LS}}^2$ is large and the estimate is pulled toward the sensing-based prior mean $\mu_{\mathrm{sen}}$. Therefore, accurate sensing information is most beneficial in pilot-limited regimes. Conversely, an inaccurate but overconfident sensing prior may degrade the estimate, which motivates evaluating mismatched-prior cases. Fig.~\ref{fig:CE} shows the channel-estimation \ac{MSE} for four cases,  indicating that more accurate sensing yields better channel estimation, with the larger gains for more noisy pilot sequences, while inaccurate sensing information introduces bias in the channel estimate.

\subsection{Impact on LLR}

After channel estimation, the received signal in \eqref{channel} can be written as
\begin{equation}
    y_{i} = \hat{h}_{\mathrm{sen}} x_{i} +\Delta \hat{h} x_{i} + w_{i}, \qquad i=1,\ldots,n, \label{channel2}
\end{equation}
where $\Delta \hat{h}\sim \mathcal{CN}(0,\sigma_{\mathrm{post}}^2)$ is the channel estimation error. Using the posterior mean $\hat{h}_{\mathrm{sen}}$, the receiver computes symbol-wise \acp{LLR} using \eqref{eq:llr_imperfect_csi}, where, for BPSK symbols with unit magnitude, the effective variance {is upper-bounded by \cite{medard2000effect}}
\begin{equation}
\sigma_{\mathrm{eff}}^2 \leq \sigma_w^2+\sigma_{\mathrm{post}}^2.
\end{equation}
This expression can be interpreted as the symbol-wise marginalization of the residual scalar channel uncertainty. Since the same channel error is common to the whole block, it induces a rank-one correlated component in the received block covariance \cite{duffy2023using}. This structure is simpler than arbitrary correlated noise, and is governed by $\sigma_{\mathrm{post}}^2$. Exploiting this correlation requires joint processing across the entire codeword. Instead, we use the corresponding symbol-wise marginal LLRs as practical reliability inputs to ORBGRAND.

Sensing information therefore benefits ORBGRAND through two mechanisms. First, a more accurate channel estimate improves the sign of the \acp{LLR}, reducing errors in the initial hard-decision vector. Second, a better uncertainty characterization improves the magnitude of the \acp{LLR}, which determines the reliability ordering used by ORBGRAND to query noise effects. As a result, accurate sensing information can reduce both the decoding error probability and the number of required noise queries, especially when the pilot budget or the ORBGRAND query budget is limited.

\section{Results}

\subsection{Simulation Scenario}

We evaluate the proposed sensing-aided ORBGRAND receiver in a SISO narrowband two-path propagation scenario. The propagation channel consists of one \ac{LoS} path and one reflected \ac{NLoS} path associated with a reflecting surface. The transmitter and receiver communicate using coded BPSK over the channel, where the channel remains constant over one coded block. The transmitter location is fixed at
$\boldsymbol{x}_{\text{Tx}}=[0,0]^{\top}\, \text{m}$. The \ac{VA} of the reflecting surface is located at
$\boldsymbol{x}_{\text{VA}}=[12,0]^{\top}\, \text{m}$, corresponding to a surface at $x=6\, \text{m}$. For each transmitted block, the receiver position is drawn uniformly from a rectangular region centered at $\boldsymbol{x}_{\text{Rx}}=[3.5,2.0]^{\top}\, \text{m}$, with a maximum displacement $1\, \text{m}$ along each coordinate. The carrier frequency is $f_{\mathrm{c}}=3.5\, \text{GHz}$. The reflection attenuation coefficient is set to $\Gamma=0.75$. The variance of the residual unmodeled channel component is set to $\sigma_{\mathrm{res}}^2=0.03$. The transmitted BPSK symbols and pilot symbols are normalized to unit energy. The noise variance is set according to the target $E_s/N_0$ as $\sigma_w^2=1/(E_s/N_0)$.
Since unit-energy pilot symbols are used, the total pilot energy is \(E_p=N_p\). We use an \ac{eBCH} $(16,11)$ code with rate $R=11/16$.  However, the instantaneous received \ac{SNR} additionally depends on the block channel gain. Unless otherwise stated, the  pilot length is set to  $N_p=2$, and the ORBGRAND's abandonment threshold is set to $T_{\max}=1000$. We compare the following receivers:
\begin{itemize}
    \item \textbf{Perfect CSI}: The receiver computes \acp{LLR} using the true channel coefficient. This serves as a benchmark.
    \item \textbf{Pilot-only}: The receiver estimates the channel using pilots only, without sensing information, corresponding to the case in Section~\ref{Sec:pilot_only}.
    \item \textbf{Rx-aided}: The receiver uses the known receiver position to construct a prior based on the \ac{LoS} component.
    \item \textbf{Rx+VA-aided}: The receiver uses both the receiver position and the \ac{VA} location to construct a two-path prior.
    \item \textbf{Mismatched prior}: The receiver uses an accurate receiver position, but an inaccurate VA location to construct the prior, which is used to assess robustness.
\end{itemize}
For each transmitted block, pilot symbols are first used to obtain an LS channel estimate. Depending on the available sensing information, this estimate is either used directly or combined with a sensing-based channel prior through LMMSE estimation. The resulting channel estimate and posterior uncertainty are then used to compute the \acp{LLR} supplied to ORBGRAND.  When the reflecting surface is not available to the receiver, the reflected component is not included in the prior mean and is instead treated as channel uncertainty. In this case, the prior variance considers the reflected-path through an additional uncertainty term $\sigma_{\mathrm{NLoS}}^2=0.25$. When both the receiver position and the \ac{VA} location are available, the reflected component is included in the prior mean, and the remaining prior variance is set to $\sigma_{\mathrm{res}}^2=0.03$. For performance evaluation, \ac{BLER} and average number of ORBGRAND queries are recorded. In total, 10000 Monte Carlo (MC) simulations are performed for each case.

\subsection{Results and Discussion}
\begin{figure}[t]
\center
\centerline{\includegraphics[width=1\linewidth]{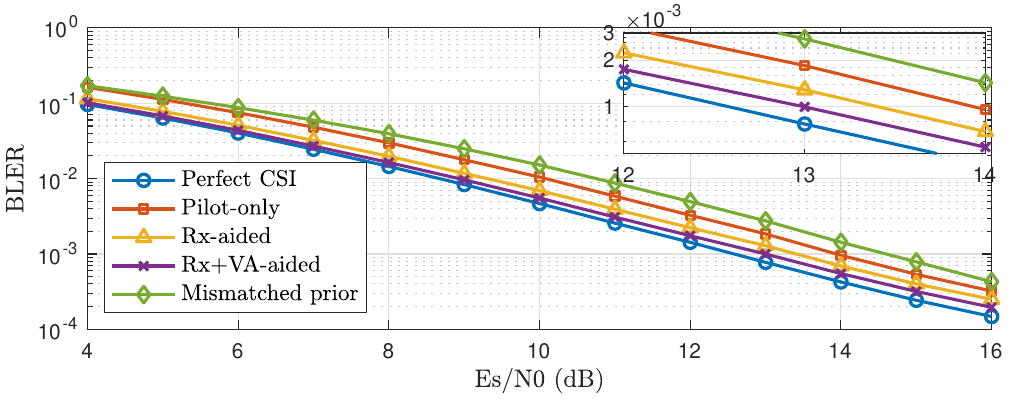}}
\vspace{-3mm}
\caption{\ac{BLER} versus $E_s/N_0$ for the five receiver configurations  ($N_p=2$).
}
\label{fig:BLER}
\end{figure}

We first evaluate the benefit of sensing information within the ORBGRAND framework. Fig.~\ref{fig:BLER} compares the \ac{BLER} of the five receiver configurations. The perfect-CSI curve is a benchmark in which the true channel coefficient is known at the receiver. In contrast, the pilot-only receiver exhibits a clear performance loss due to channel-estimation error, especially at low and medium $E_s/N_0$. Incorporating sensing information improves performance: the Rx-aided receiver leverages the known receiver position to form a prior for the \ac{LoS} component, yielding a more accurate channel estimate and, hence, more reliable \acp{LLR}. The Rx+VA-aided receiver achieves the best practical performance and closely approaches the perfect-CSI benchmark across the simulated $E_s/N_0$ range, providing an approximately 1.8~dB gain over the pilot-only case. This indicates that, when both receiver position and reflector information are available, the resulting two-path prior significantly improves the channel estimate and thus the \acp{LLR} supplied to ORBGRAND. Conversely, the mismatched-prior case can underperform even the pilot-only receiver, especially at higher $E_s/N_0$, demonstrating that inaccurate environmental information may degrade decoding when used with high confidence. At low $E_s/N_0$, the LS estimation error dominates and partially masks the bias introduced by the mismatched prior, leading to a small apparent difference.

The same ordering observed in \ac{BLER} is reflected in the query complexity: the pilot-only receiver requires more queries because channel-estimation errors perturb both the \ac{LLR} signs and their reliability magnitudes, causing ORBGRAND to explore less favorable noise effects. Sensing-aided receivers reduce the query count. At $E_s/N_0=4$~dB, the Rx-aided and Rx+VA-aided receivers reduce the average query count by 13.2\% and 18.2\%, respectively, compared to the pilot-only receiver, while the perfect-CSI receiver achieves a 20.0\% reduction. The performance gap narrows at high $E_s/N_0$, where the LMMSE estimator relies more on pilot observations and the residual error approaches that of LS.  Overall, sensing improves ORBGRAND performance not by modifying the decoder, but by improving the reliability ordering of \acp{LLR} via more accurate and precise channel estimates, which enhances both decoding accuracy and query efficiency.

\begin{figure}
\center
\centerline{\includegraphics[width=1\linewidth]{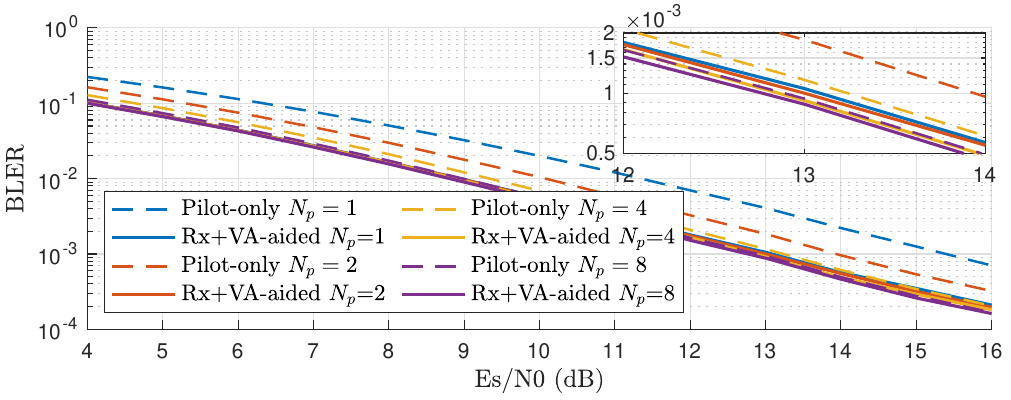}}
\vspace{-3mm}
\caption{\ac{BLER} versus $E_s/N_0$ for the pilot-only and the Rx+VA-aided receivers using different $N_p$.
}
\label{fig:BLERPILOT}
\end{figure}

Fig.~\ref{fig:BLERPILOT} evaluates the impact of pilot length. As the number of pilot symbols increases, the pilot-only receiver improves because the LS estimation variance decreases with pilot energy. Similar trends are observed for the sensing-aided receivers; however, the improvement is less pronounced because the prior information already reduces the estimation uncertainty, so the remaining error does not decrease as rapidly with pilot length. The Rx+VA-aided receiver remains markedly better in the pilot-limited regime: with very short pilots, the geometry-based prior compensates for scarce pilot observations by providing a good prediction of the channel coefficient. This suggests that sensing information can reduce the pilot overhead required to achieve a target \ac{BLER}. As the pilot length increases, the gap between the pilot-only and the Rx+VA-aided receivers narrows, since the pilot-only channel estimate becomes more accurate and there is less room for the prior to improve the estimate. {This experiment keeps $E_s/N_0$ and uses unit-energy pilots, so increasing $N_p$ also increases the total pilot energy. Thus, Fig.~\ref{fig:BLERPILOT} shows the benefit of additional pilot resources rather than an energy-fair comparison. With fixed total block energy, longer pilots would trade improved channel estimation against reduced data-symbol energy.} Since the Rx+VA-aided receiver achieves a low \ac{BLER} with much shorter pilot sequences, it would provide even larger gains in such an energy-constrained scenario, where the pilot-only receiver would require longer pilots to achieve comparable performance.

\section{Conclusions}

This paper investigated sensing-aided ORBGRAND over a SISO narrowband fading channel. Environmental information was incorporated through a channel prior and fused with pilot observations via LMMSE estimation. The resulting posterior mean and uncertainty were then used to compute the \acp{LLR} supplied to ORBGRAND. Simulation results showed that sensing-aided estimation improves the \ac{BLER} and reduces the average number of ORBGRAND queries, with the largest gains occurring in pilot-limited regimes. These gains arise from the improved  reliability ordering of the \acp{LLR} enabled by more accurate and precise channel estimates. We also observed that mismatched environmental priors can degrade performance, underscoring the importance of accurate sensing information and uncertainty-aware prior modeling. Future work will explicitly account for sensing uncertainty and exploit block-level channel correlation. 

\scriptsize{
\section*{Acknowledgment}
This work was partially supported by the Wallenberg AI, Autonomous Systems and Software Program (WASP) funded by Knut and Alice Wallenberg Foundation, and by the National Science Foundation under Grant No. (NSF grant number ECCS-2433994 and ECCS-2433996).}
\bibliography{IEEEabrv,Bibliography}

\end{document}